\documentclass[prx,reprint,citeautoscript,groupedaddress]{revtex4-1}
\usepackage{amsmath}
\usepackage{bm}
\usepackage{graphicx,color}
\usepackage{stix}
\usepackage{subfigure}
\usepackage[usenames,dvipsnames]{xcolor}

\newcommand*{\abinitio}{{\textit{ab initio}\,}}
\newcommand*{\GW}{{\textit{GW}}}
\newcommand*{\gw}{{\textit{G}\textsubscript{0}\textit{W}\textsubscript{0}}}
\newcommand*{\hematite}{{$\alpha$-Fe\textsubscript{2}O\textsubscript{3}}}
\newcommand*{\feo}{{Fe\textsubscript{2}O\textsubscript{3}}}
\newcommand*{\srmo}{{SrMoO\textsubscript{3}}}

\newcommand{\eri}[2]{{\left( #1 \middle| #2 \right)}}

\newcommand*{\veck}{{\mathbf{k}}}

\newcommand*{\vecG}{{\mathbf{G}}}

\newcommand*{\vech}{{\mathbf{h}}}
\newcommand*{\vecF}{{\mathbf{F}}}
\newcommand*{\vecSig}{{\mathbf{\Sigma}}}
\newcommand*{\vecDel}{{\mathbf{\Delta}}}

\newcommand*{\vecI}{{\mathbf{I}}}

\newcommand*{\vecPi}{{\mathbf{\Pi}}}

\newcommand*{\emb}{{\mathrm{emb}}}

\newcommand*{\imp}{\mathrm{imp}}

\newcommand*{\occ}{{\mathrm{occ}}}

\newcommand*{\vir}{{\mathrm{vir}}}

\newcommand*{\AO}{{\rm AO}}

\newcommand*{\DC}{{\rm DC}}

\definecolor{myblue}{rgb}{0,0,1}
\usepackage[breaklinks=true,colorlinks=true,linkcolor=myblue,urlcolor=myblue,citecolor=myblue]{hyperref}

\begin{document}
\title{\textit{Ab initio} Full Cell \GW+DMFT for Correlated Materials}
\author{Tianyu Zhu}
\email{tyzhu@caltech.edu}
\author{Garnet Kin-Lic Chan}
\email{gkc1000@gmail.com}
\affiliation{Division of Chemistry and Chemical Engineering, California Institute of Technology, Pasadena CA 91125}

\begin{abstract}
  The quantitative prediction of electronic properties in correlated materials requires simulations
  without empirical truncations and parameters.
  We present a method to achieve this goal through a new \abinitio formulation of dynamical mean-field theory (DMFT).
  Instead of using small impurities defined in a low-energy subspace, which require 
  complicated downfolded interactions which are often approximated, 
  we describe a full cell \GW+DMFT approach, where the impurities comprise all atoms in a unit cell or supercell
  of the crystal. Our formulation results in large impurity problems,
  which we treat here with efficient quantum chemistry impurity solvers that work on the
  real-frequency axis, combined with a one-shot \gw~treatment of long-range interactions.
  We apply our full cell approach to bulk Si, two antiferromagnetic correlated insulators NiO and \hematite, and the paramagnetic correlated metal \srmo, with impurities containing up to
  10 atoms and 124 orbitals. We find that spectral properties, magnetic moments, and two-particle spin correlation functions are obtained in good agreement with experiment. In addition, in the metal oxide insulators, the balanced treatment of correlations involving all orbitals in the cell leads
  to new insights into the orbital character around the insulating gap.

\end{abstract}

\maketitle

\section{Introduction}
Computing the properties of correlated electron materials with quantitative accuracy remains a fundamental challenge in
\abinitio simulations~\cite{Kent2018a}. This is because strong electron interactions, for example in materials with open $d$ or $f$ shells,
can lead to emergent phases such as high-temperature superconductivity, which cannot be described by the mean-field and low-order perturbation
approximations commonly employed by \abinitio methods.

Quantum embedding methods~\cite{Georges1992,Georges1996,Knizia2012a,Sun2016} in principle provide a promising
route to access the phase diagrams of correlated materials, because they simultaneously treat strong local electron interactions and the thermodynamic limit.
 Among the different variants of quantum embedding used for this purpose,
the combination of dynamical mean-field theory (DMFT) (and
its cluster extensions~\cite{Kotliar2001,Hettler2000}) and density functional theory (DFT)~\cite{Kohn1965}, known as DFT+DMFT, is very popular~\cite{Held2006,Kotliar2006,Held2007}.
In this combination, one  views DFT as a low-level theory that accounts for band structure and the long-range interactions,
while the high-level solution of the DMFT impurity problem, defined on a small set of correlated orbitals, introduces
diagrams arising from the strong local interactions.
Yet despite many successes, DFT+DMFT does not  provide a truly parameter-free and quantitative \abinitio theory of correlated materials, due to two
closely related issues. First, the local Coulomb interaction in the DMFT impurity problem is typically treated as an adjustable Hubbard-like parameter~\cite{Nilsson2018}, or is else estimated within another approximation~\cite{Aryasetiawan2004}. Second, a double-counting correction~\cite{Wang2012,Haule2015} is required to remove the DFT contribution to the local interactions, but no consistently accurate double-counting correction is known~\cite{Karolak2010}. Beyond these two primary concerns arising from the local interactions, density functionals also do not always reliably account for the long-range interaction effects~\cite{Perdew2017}.

To obtain a truly quantitative, \abinitio formulation of DMFT, one must work within a diagrammatically clean formalism.
In this context, it is natural to replace DFT with the \GW~approximation~\cite{Hedin1965a} as the low-level theory.
The \GW~approximation (often employed in its one-shot form (\gw))~\cite{Hybertsen1986,Shishkin2006} has been shown to fix
many of the problems with semilocal density functionals (such as the underestimation of band gaps of weakly-correlated semiconductors) and
thus appears a practical way to include the most important low-order long-range interaction diagrams. 
The combination with DMFT can then be formulated without double-counting
by exactly subtracting the local \GW~contributions. The idea of self-consistently embedding the impurity self-energy and
contributions to the polarization propagator arising
from long-range interactions was proposed almost 18 years ago as the \GW+(E)DMFT approximation~\cite{Sun2002,Biermann2003}, but only very recently have
self-consistent implementations appeared~\cite{Boehnke2016,Nilsson2017}. However, while these developments are promising, applications have remained more limited than those with DFT+DMFT and have retained some problematic issues of that approach~\cite{Tomczak2012,Taranto2013,Tomczak2017,Choi2016a,Choi2019a,Petocchi2020}.
In particular, all current \GW+(E)DMFT methods still require strongly downfolded interactions, because the impurity is restricted to the truncated low-energy subspace of a few correlated $d$ or $f$ orbitals (Fig.~\ref{fig:illustration}(b)). Downfolding to a small number of strongly coupled orbitals is numerically challenging, and yields retarded interactions that either limit the applicable impurity solvers or
which must be truncated or otherwise approximated.
If one ignores the embedding of the polarization propagator to work purely with the bare interactions, one obtains
the self-energy embedding theory (SEET)~\cite{Lan2017}. 
However, applications of this simpler approach in realistic solids have only appeared very recently~\cite{Iskakov2020}. Aside from these technical issues, in some more complex correlated materials, the local orbitals can be intertwined with
other itinerant bands~\cite{Lu2008,Weber2010}. In such cases, even defining a set of local correlated orbitals can be difficult, and the quality
of the calculation then depends sensitively on this choice~\cite{Amadon2008}.


A common origin of many of the above challenges is the definition of the impurity problem in terms
of a small low-energy subspace. This is done only to obtain as simple an
impurity problem as possible, as motivated by model Hamiltonians, but it is not
a requirement of the more general DMFT formalism.
Consequently, in this work, we present a new formulation, which we term
\abinitio full cell \GW+DMFT. 
In this approach we define the impurity to be the full unit cell - or even multiple unit cells of atoms - where
each atom is described by a large localized set of atomic orbitals, covering the core, valence and high-energy virtual orbitals (Fig.~\ref{fig:illustration}(a)). Since no low-energy subspace is identified, there is no downfolding and its associated uncertainties, and we can simply use the full set of
bare Coulomb interactions between the impurity orbitals,
avoiding theoretical and numerical ambiguities.
(A related full cell idea has been used to enable \GW-in-DFT Green's function embedding~\cite{Chibani2016}).
In addition, important medium range (non-local) interaction effects, weaker than the strongest local correlations, but stronger
  than the interactions captured by \GW, can now enter the theory. (Such interactions, which are neglected in standard \GW+DMFT treatments,
  are known to yield significant errors
  in a variety of settings, for example in the treatment of metallic sodium~\cite{Nilsson2017}).
While conceptually simple, our full cell approach engenders
two new technical complexities.
The first is the need to set up the large impurity problem (for example to
efficiently generate all the matrix elements) but this is enabled by technical advances we have made in 
the PySCF simulation platform~\cite{Sun2018} and our recently developed general \abinitio quantum embedding framework~\cite{Cui2020,Zhu2020}.
The second is the need to solve the resulting impurity problem with a large number of orbitals.
Here, the key insight is that many orbitals in the full cell impurity are only weakly correlated, and impurity solvers which take
advantage of this, such as those used in molecular quantum chemistry~\cite{Zgid2011,Zgid2012} can then work very efficiently.
In this work, we will use two such kinds of solvers: a coupled-cluster singles and doubles (CCSD) Green's function solver~\cite{Zhu2019,Shee2019} as well as a selected configuration interaction (selected CI) solver~\cite{Holmes2016}, carrying out self-consistency along the real-frequency axis.
The accuracy of such solvers in impurity problems has been benchmarked elsewhere~\cite{Zgid2012, lu2014efficient, Go2017, Zhu2019}:
thus we emphasize that a specific choice of the solver is not the message of this work. Rather, our 
focus is on establishing a generic full cell embedding framework that avoids downfolding and introduces new non-local physics at
two levels of approximation at medium and large distances, and whose mathematical formulation allows
a natural combination with a wide variety of quantum chemistry impurity solvers, including the two used here.
We apply the full cell \GW+DMFT method to compute the spectral properties of Si, the spectra, magnetic moments and spin correlation functions of two correlated insulators, NiO and \hematite, in their antiferromagnetic (AFM) phases, as well as the spectra of a paramagnetic correlated metal \srmo. 
Our largest calculation in hematite uses an impurity of four Fe and six O atoms, giving rise to an unprecedentedly large \abinitio DMFT impurity problem
with 124 impurity orbitals. 

\section{Theory and implementation}

In the full cell \GW+DMFT formulation, because the impurity cell  contains all atoms in a crystal cell (or, more usually, a supercell),
the effects of what would normally be thought of as long-range interactions
on the polarization and self-energy from within the supercell are all included.
Also, because the supercell is treated at a level beyond \GW, we include the medium-range
  interactions that are neglected in standard \GW+DMFT treatments.
However, we will treat contributions from long-range interactions beyond the supercell
only at the level of the self-energy matrix of the crystal, computed at the one-shot \gw~level.
Because of this, certain contributions to the polarization propagator involving interactions far from the supercell,
that would require the bosonic self-consistency of EDMFT~\cite{Nilsson2017}, are omitted. 
In practice, this means that rather than partitioning into the strictly strongest interactions (for example in a $d$- or $f$-shell), coupled to a low-level treatment of everything else as in standard \GW+DMFT, our formulation prioritizes a higher-level treatment of interactions within a significant distance of the most correlated orbitals. 
  The magnitude of the neglected effects from the lack of bosonic self-consistency, while ameliorated
  by the explicit treatment of large cells, can only be established
  through numerical experiments, as performed below.
(An example of a case where the lack of bosonic self-consistency could be qualitatively important
  is in systems where there is an overlap of plasmon satellites and Hubbard bands~\cite{Boehnke2016}).


Given a periodic crystal, we start by performing a one-shot \gw~calculation on top of a mean-field reference (DFT or HF), using crystalline Gaussian atomic orbitals and Gaussian density fitting (GDF) integrals~\cite{Sun17}. Because the \gw~approximation is reference dependent, we will
denote the approximation \gw@\text{reference}.
The full \gw~self-energy matrix is computed in the mean-field molecular orbital (MO) basis along the imaginary-frequency axis~\cite{Ren2012,Wilhelm2016}:
\begin{equation}\label{eq:g0w0}
\begin{split}
\vecSig_{nn'}^{\GW} (\veck, i\omega) = & -\frac{1}{2\pi} \sum_{m\veck_m} \int_{-\infty}^{\infty} d\omega' [\vecG_0(\veck_m, i\omega-i\omega')]_{mm}  \\
&  \times \sum_{PQ\veck_P} v_P^{nm} [\vecI-\vecPi(\veck_P, i\omega')]_{PQ}^{-1} v_Q^{mn'},
\end{split}
\end{equation}
where $v_P^{nm}$ represents the 3-index electron repulsion integral (ERI) $\eri{P\veck_P}{n\veck_n m\veck_m}$, $P$ is the Gaussian auxiliary basis, and $n$ and $m$ represent mean-field molecular orbitals (bands). $\veck_P$, $\veck_n$ and $\veck_m$ satisfy crystal momentum conservation: $\veck_P = \veck_n - \veck_m + n\mathbf{b}$, where $\mathbf{b}$ is a lattice vector, and $\vecG_0(\veck_m, i\omega-i\omega')$ is the mean-field Green's function. The integration in Eq.~\ref{eq:g0w0} is carried out efficiently using a modified Gauss-Legendre grid~\cite{Ren2012} (100 grid points were used in this study). The polarization kernel $\vecPi(\veck_P, i\omega')$ is 
\begin{equation}\label{eq:Pi}
\begin{split}
\vecPi_{PQ}(\veck_P, i\omega') = 2\sum_{i\veck_i}^{\occ}\sum_{a\veck_a}^{\vir} v_{P}^{ia} \frac{\epsilon_{i\veck_i} - \epsilon_{a\veck_a}}{\omega'^2 + (\epsilon_{i\veck_i} - \epsilon_{a\veck_a})^2} v_{Q}^{ai} ,
\end{split}
\end{equation}
where $\epsilon_{i\veck_i}$ and $\epsilon_{a\veck_a}$ are occupied and virtual orbital energies respectively. Note that
in a Gaussian basis formulation, the number of bands and size of
auxiliary basis are significantly smaller than in plane-wave \GW~formulations~\cite{Booth2016}, and
because of this, the summations in Eqs.~\ref{eq:g0w0}-\ref{eq:Pi} run over all bands. To obtain the real-frequency \gw~self-energy,
we perform analytic continuation. Here, we fit the self-energy matrix elements to $N$-point Pad\'e approximants ($N=18$ in this work) using Thiele's reciprocal difference method~\cite{Vidberg1977}. For the detailed \gw~algorithm, we refer readers to Ref.~\cite{Zhu2020b}.
Our \gw~scheme thus allows us to use  continuation to obtain the retarded self-energy on the real axis at arbitrary broadenings, although
as an alternative, a contour deformation scheme~\cite{Godby1988} may also be used to directly compute the real-frequency time-ordered self-energy
without the need for analytic continuation. This latter strategy may be attractive to explore in the future to completely avoid possibly unstable analytic continuations.

To define the impurity problem, we first construct an orthogonal atom-centered local orbital (LO) basis. As in our previous work on \abinitio HF+DMFT and density matrix embedding theory (DMET), we employ crystalline intrinsic atomic orbitals (IAOs) and projected atomic orbitals (PAOs) as the local orthogonal basis~\cite{Knizia2013d,Cui2020,Zhu2020}. IAOs are a set of valence atomic-like orbitals that exactly span the occupied space of the mean-field calculations, whose construction only requires projecting the DFT/HF orbitals onto predefined valence (minimal) AOs. PAOs, on the other hand, provide the remaining high-energy virtual atomic-like orbitals that complete the atomic basis and capture the correlation and screening effects.


\begin{figure}[hbt]
\centering
\includegraphics[width=3.375in]{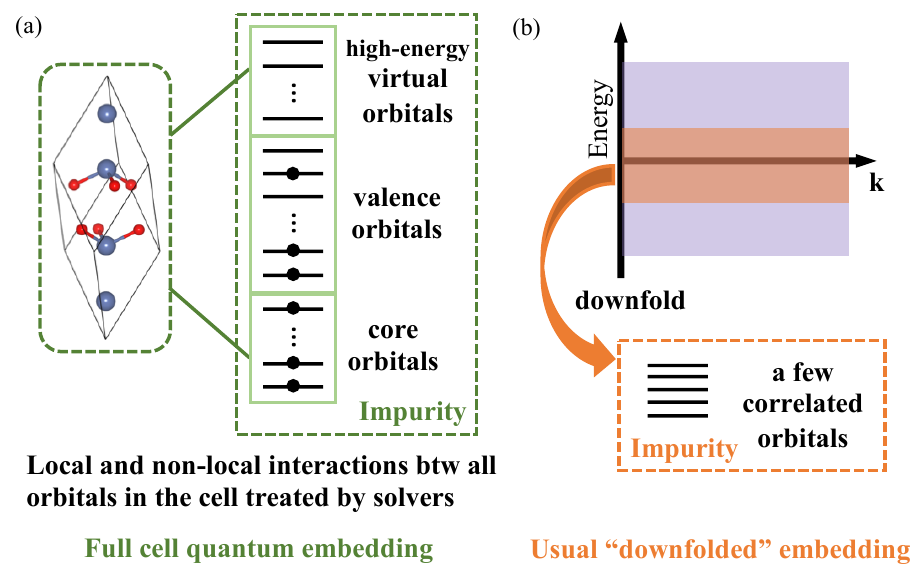}
\caption{
 Illustration of (a) \abinitio full cell \GW+DMFT and (b) usual \GW+DMFT schemes. In full cell embedding, all orbitals in the full unit cell (four Fe and six O atoms) are taken as the impurity in the \hematite~calculation. In contrast, other \GW+DMFT formulations define
  the impurity problem to contain a few correlated orbitals within a low-energy subspace
  which interact via downfolded, retarded, interactions.}
\label{fig:illustration}
\end{figure}

The impurity consists of all LOs (i.e. all IAOs and PAOs) within the impurity cell (crystal cell or supercell) with IAOs representing the core and valence orbitals and PAOs representing the high-energy
virtual orbitals. This is illustrated in Fig.~\ref{fig:illustration}(a). The most expensive step in forming the impurity Hamiltonian is  computing the bare Coulomb interaction matrix $\eri{ij}{kl}$ for all orbitals within the impurity cell. However, using Gaussian density fitting, we
can do this at relatively low cost (scaling asymptotically as  $\mathcal{O}(N_{\veck}^2N_LN_{\AO}^3)$, where $N_{\veck}$, $N_L$ and $N_{\AO}$ are the numbers of $\veck$ points and auxiliary Gaussian and atomic orbitals within the impurity cell). We refer readers to Ref.~\cite{Zhu2020} for a detailed algorithm. The impurity Hamiltonian (without bath orbitals) is therefore
\begin{equation}\label{eq:Himp}
  \hat{H}_\imp = \sum_{ij \in \imp} {\tilde{F}}_{ij} {a}^\dag_{i} {a}_{j} \\
  + \frac{1}{2} \sum_{ijkl \in \imp} \eri{ij}{kl} {a}^\dag_{i} {a}^\dag_{k} {a}_{l} {a}_{j} ,
\end{equation}
with the one-particle interaction $\tilde{F}_{ij}$ defined as the Hartree-Fock effective Hamiltonian with the double-counting term subtracted
\begin{equation}
\tilde{F}_{ij} = (F_\imp)_{ij} - \sum_{kl \in \imp} (\gamma_\imp)_{kl} [\eri{ij}{lk} - \frac{1}{2} \eri{ik}{lj}] ,
\end{equation}
and with $\gamma_\imp$ as the impurity block of the mean-field density matrix.

We then start the DMFT cycle with an initial guess of the impurity self-energy as the \gw~local 
self-energy: $\vecSig_\imp(\omega) = \vecSig_{\DC}^\GW(\omega)$. The \gw~local 
self-energy is computed in the LO basis within the impurity cell:
\begin{equation}\label{eq:gwdc}
\begin{split}
[\vecSig_{\DC}^{\GW}(i\omega)]_{ij} = & -\frac{1}{2\pi} \sum_{kl} \int_{-\infty}^{\infty} d\omega' [\vecG_0^\imp(i\omega-i\omega')]_{kl}  \\
&  \times \sum_{RS} L_R^{ik} [\vecI-\vecPi(i\omega')]_{RS}^{-1} L_S^{lj},
\end{split}
\end{equation}
and analytically continued to the real axis. Here, all local orbital indices ($i,j,k,l$) are within the impurity cell. The 3-index tensor $L_R^{ij}$ is
computed from a Cholesky decomposition of the impurity ERI: $\eri{ij}{kl} = \sum_{R} L_R^{ij} L_R^{kl}$. Note that the polarization
propagator is computed in the impurity orbital space, first in the imaginary time domain~\cite{Liu2016,Wilhelm2018}:
\begin{equation}\label{eq:Pidc}
\vecPi_{RS}(\tau) = \sum_{ijkl \in \imp} L_P^{ij} [\vecG_0^\imp (\tau)]_{ki} L_Q^{kl} [\vecG_0^\imp (-\tau)]_{lj} ,
\end{equation}
and then cosine transformed into imaginary frequency space.

The hybridization self-energy is then computed:
\begin{equation}\label{eq:hyb}
\vecDel (\omega) = (\omega + \mu) \vecI - \tilde{\vecF}  -\vecSig_{\imp}(\omega) - \vecG^{-1}(\mathbf{R}=\mathbf{0},\omega) ,
\end{equation}
with the lattice Green's function defined as
\begin{equation}\label{eq:Gloc}
\vecG (\mathbf{R}=\mathbf{0},\omega) = \frac{1}{N_{\veck}} \sum_{\veck} [(\omega + \mu) \vecI -\vech(\veck)  -\vecSig(\veck, \omega)]^{-1} ,
\end{equation}
and the full \GW+DMFT self-energy defined as
\begin{equation}\label{eq:sigmak}
\vecSig(\veck, \omega) = \vecSig^\GW(\veck, \omega) + \vecSig_\imp(\omega) - \vecSig_\DC^\GW(\omega).
\end{equation}
Here, $\mu$ is the chemical potential, which is adjusted during the DMFT self-consistency to ensure that the electron count of the impurity is correct. $\vech(\veck)$ is the bare one-particle Hamiltonian for the whole solid. We subtract the local \gw~self-energy in Eq.~\ref{eq:sigmak}, and the DFT exchange-correlation potential is excluded from both the impurity and lattice self-energies, consequently our method exactly avoids double-counting. When the one-shot \gw~is used, it is not  guaranteed that the \GW+DMFT self-energy in Eq.~\ref{eq:sigmak} is strictly causal, and a fully self-consistent \GW+DMFT is formally required~\cite{Lee2017a}. However, in our test cases, we have not observed non-causal negative spectral functions at low-energies. Because multiple orbitals and atoms are now chosen as the impurity, the final \GW+DMFT self-energy does not strictly preserve all the symmetries of the solid, similar to other cellular DMFT methods~\cite{Kotliar2001}. Possible ways to alleviate this problem include incorporating more cells as the impurity, or deriving a translational and crystal symmetry invariant impurity Hamiltonian,
  as done in the dynamical cluster approximation (DCA)~\cite{Hettler2000}.

In order to use a wavefunction (Hamiltonian-)based impurity solver, we discretize $\vecDel (\omega)$. We  discretize along the real-frequency axis~\cite{DeVega2015} so that dynamical quantities (e.g., spectral functions) are obtained more accurately.
To obtain the discretization, we start with an initial guess by direct discretization on a numerical grid and optimize bath couplings $\{ V_{ip}^{(n)} \}$ and energies $\{ \epsilon_n \}$ to minimize a cost function
over a range of real-frequency points (see Supplemental Material):
\begin{equation}\label{eq:bathfit}
D = \sum_{\omega_l} \sum_{ij} \Big( \Delta_{ij}(\omega_l+i\eta) - \sum_{n=1}^{N_\epsilon} \sum_{p=1}^{N_p} \frac{V_{ip}^{(n)} V_{jp}^{(n)}}{\omega_l +i\eta - \epsilon_n} \Big) ^2,
\end{equation}
where $N_\epsilon$ is the number of bath energies and $N_p$ is the number of bath orbitals per bath energy, and we
use a  broadening factor $\eta=0.1$ a.u. unless specified. This optimization scheme is necessary to reduce the number of bath orbitals and avoid non-causal behavior in the \GW+DMFT self-energies. The bath degrees of freedom are truncated by only coupling bath orbitals to the valence IAOs, further reducing  computational and optimization costs. The full embedding problem with both impurity and bath orbitals is thus defined from the Hamiltonian
\begin{equation}\label{eq:Hemb}
  \hat{H}_\emb = \hat{H}_\imp + \sum_{n=1}^{N_\epsilon} \sum_{p=1}^{N_p} \Big( \sum_i V_{ip}^{(n)} (a_i^\dag a_{np} + a_{np}^\dag a_i) + \epsilon_n a_{np}^\dag a_{np} \Big) .
\end{equation}

We solve for the ground-state and Green's functions of the impurity Hamiltonian using
two quantum chemistry impurity solvers. For completeness, we give some general background on the solvers. 
The first is a CCSD Green's function solver at zero temperature.
 Our implementation
 of the CCSD Green's function solver is able to treat around 200 (impurity + bath) orbitals.
  At the singles and doubles level,
  CC may be viewed as generating ring, ladder, and coupled ring-ladder diagrams, and is exact for (arbitrary products of) two-electron problems regardless of interaction strength. CCSD is based on time-ordered diagrams, and the corresponding CCSD Green's function does not include contributions from all time-orderings of the ring diagrams that come from the \GW~self-energy, but contains a large number of vertex corrections to the self-energy and polarization propogator~\cite{Lange2018a}.
It has been shown to be accurate in a variety of settings,
including simple metallic and ordered magnetic states in \abinitio calculations~\cite{McClain2016,Gao2020,Williams2020},  across weak to strong couplings
when employed with small cluster DMFT impurities in Hubbard-like models~\cite{Zhu2019}, and in electron gases up to moderately dilute densities, e.g. that
of metallic sodium~\cite{McClain2016}. Standard implementations of CCSD are capable of treating general Coulomb interactions and hybridizations, and recent work has demonstrated \abinitio calculations in materials at finite temperatures~\cite{White2018a, White2020}.
Similar to other quantum chemistry based approaches, the efficiency of the CCSD method stems from 
taking advantage of the fact that many orbitals (e.g., far from the Fermi surface) are nearly full or empty. However, it is less suited to spin-fluctuations, and because CCSD truncates the
coupled-cluster excitations at low order, it will break down when large spin fluctuations connecting many electrons
simultaneously appear. In practice, a key indicator for the breakdown of CCSD is the magnitude of its excitation amplitudes, with large amplitudes suggesting inaccurate results.

In cases where CCSD breaks down, one can use higher excitation levels in CC~\cite{McClain2016}, or other impurity solvers can be employed.
Other interesting quantum chemistry solvers in this context include  
quantum chemistry DMRG (QC-DMRG)~\cite{Chan2011} and selected configuration interaction (selected CI)~\cite{Holmes2016,Huron1973,Mejuto-Zaera2019,Zgid2012}
methods. For example, QC-DMRG offers a robust way to obtain correlated ground-states in complex systems~\cite{li2019electronic} and \abinitio Green's functions with up to 50 or more strongly correlated orbitals~\cite{Ronca2017a}. 
  Selected CI methods are based on an excitation picture, and thus are very efficient with
  large numbers of nearly filled and empty orbitals ($\sim$200), although they are limited to smaller numbers
  of  strongly correlated orbitals than DMRG ($\sim$20). These two solvers are systematically improvable towards numerical exactness, by decreasing the selection threshold (selected CI) or increasing the bond dimension (DMRG). To illustrate the generality of our embedding framework as well as
  to benchmark the accuracy of the CCSD solver, we show results also from a selected CI impurity solver (based on the semistochastic heat-bath configuration interaction (SHCI) method~\cite{Holmes2016,Sharma2017,Li2018,Yao2020}) within our \GW+DMFT approach and apply it to the correlated metal \srmo.

From the impurity Green's functions computed by the CCSD or SHCI solvers, we obtain an updated impurity self-energy $\vecSig_\imp(\omega)$, and from this the DMFT cycle (Eqs.~\ref{eq:hyb}-\ref{eq:Hemb}) is iterated until convergence between the impurity and lattice Green's functions:
\begin{equation}
\vecG_{\imp} (\omega) = \vecG(\mathbf{R}=\mathbf{0},\omega).
\end{equation}

\section{Results}

We first apply our method to crystalline silicon. Although Si is considered a weakly-correlated semiconductor, it is
still a challenging system for many DFT functionals (such as LDA and GGA) which do not yield accurate band gaps~\cite{Heyd2005}.  One-shot \gw~on top of LDA or GGA
is known to significantly improve the band structure, although this relies somewhat on the cancellation of errors~\cite{Shishkin2007a}.
Such a small band-gap system also poses challenges to quantum embedding methods that start from a local correlation picture, such as DMFT~\cite{Zein2006}, due to the long-range nature of its statically screened Coulomb interaction, which must be included in the treatment.

\begin{figure}[hbt]
\centering
\subfigure{
\includegraphics{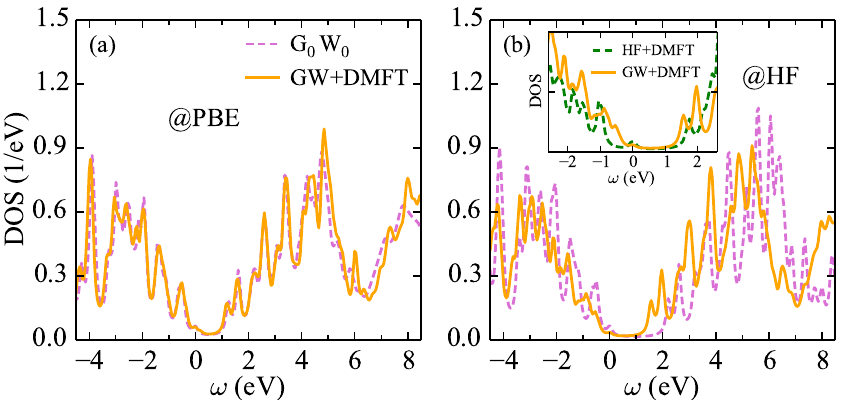}}
\vfill
\subfigure{
\includegraphics{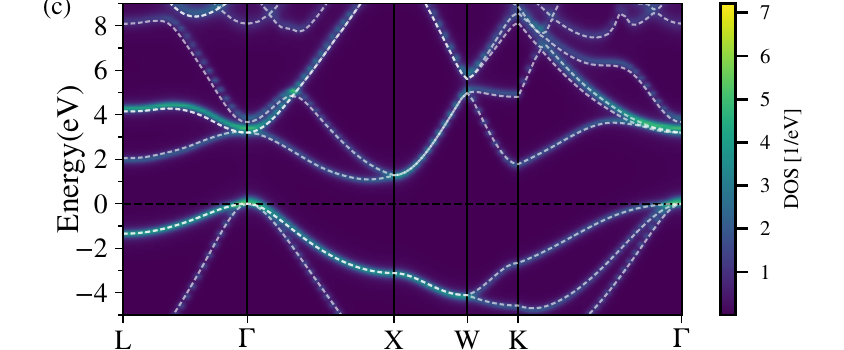}}%
\caption{Full cell \GW+DMFT results for silicon. (a)(b) Local spectral function from \GW+DMFT (PBE reference) and \GW+DMFT (HF reference). Inset of (b): \GW+DMFT compared to HF+DMFT ($4\times4\times4$ $\veck$-mesh and GTH-DZVP basis) DOS taken from Ref.~\cite{Zhu2020}. (c) Band structure
  starting from PBE orbitals. The heat map represents the \GW+DMFT@PBE result and the dashed line gives the \gw~bands. A broadening factor of 0.1 eV is used.}
\label{fig:si}
\end{figure}

The full cell \GW+DMFT results for Si are presented in Fig.~\ref{fig:si}. We used the GTH-PADE pseudopotential~\cite{Hartwigsen98} and GTH-TZVP basis~\cite{Vandevondele2005}, and a $6\times6\times6$ $\Gamma$-centered $\veck$-point sampling. The impurity was defined as the unit cell of 2 Si atoms with 34 local orbitals ($3s3p3d4s4p5s5p$ for Si), and 128 bath orbitals were used. As known from other \gw~calculations~\cite{Fuchs2007} and as seen in Figs.~\ref{fig:si}(a) and~\ref{fig:si}(b), the mean-field starting point strongly affects the quality of the \gw~results; \gw@PBE gives an accurate band gap of 1.09 eV when compared to the experimental value of 1.17 eV~\cite{ODonnell1991}, while \gw@HF overestimates the band gap, giving 2.04 eV. \GW+DMFT predicts the band gap of Si to be 1.01 eV (@PBE) and 1.39 eV (@HF), largely removing the reference dependence of \gw, due to the more complete
inclusion of diagrams from interactions within the unit cell.
The spectral function is also greatly improved in \GW+DMFT compared to \gw@HF. In earlier HF+DMFT calculations~\cite{Zhu2020} (see inset of Fig.~\ref{fig:si}(b)), we found the band gap
to be too large by 0.5 eV, and this quantifies the effect of the long-range correlations in \gw~on the band gap of Si. From Fig.~\ref{fig:si}(c), we note that \GW+DMFT@PBE maintains the accurate band structure of \gw@PBE, in contrast to self-consistent \GW,~which is known to
lead to worse results than \gw~itself~\cite{Shishkin2007,Kutepov2017}.

\begin{figure}[hbt]
\centering
\includegraphics{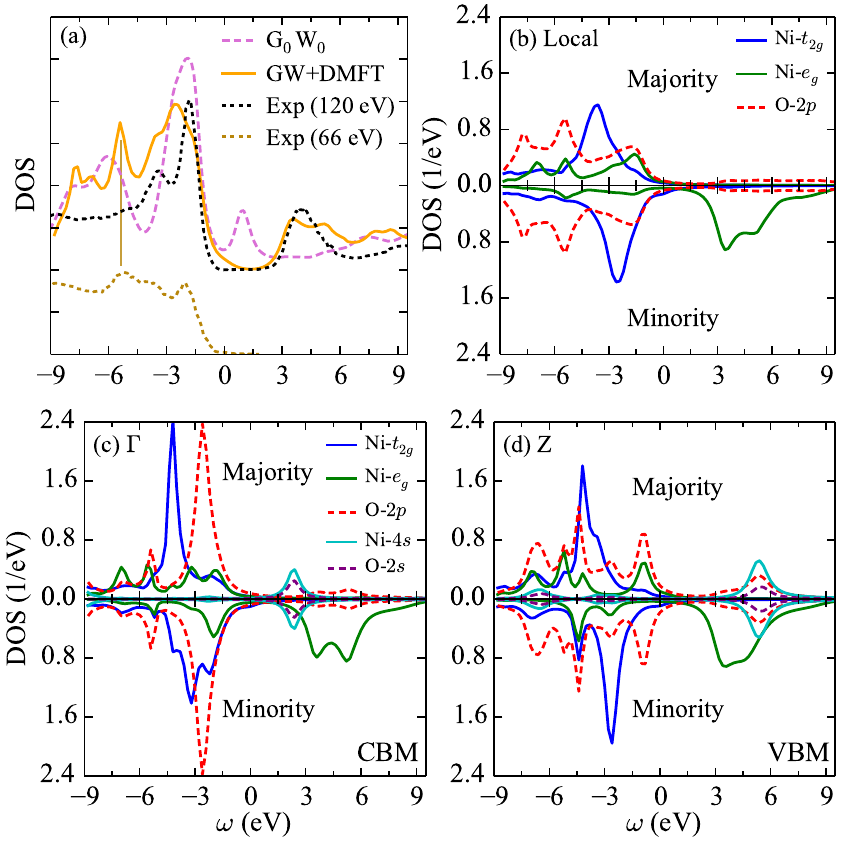}
\caption{Full cell \GW+DMFT results for NiO (AFM phase) based on the PBE reference. (a) Local DOS. The experimental spectra are taken from Ref.~\cite{Sawatzky1984}. (b) Orbital-resolved local DOS. (c)(d) Orbital-resolved and momentum-resolved DOS at the $\Gamma$ point (CBM) and $\mathrm{Z}=(0.5,0.5,0.5)$ point (VBM). A broadening factor of 0.4 eV is used.}
\label{fig:nio}
\end{figure}

We next show the results of full cell \GW+DMFT in Fig.~\ref{fig:nio} for a strongly-correlated insulator, NiO, in the antiferromagnetic phase. The GTH-PADE pseudopotential and GTH-DZVP-MOLOPT-SR basis set~\cite{VandeVondele2007} were used with a $6\times6\times6$ $\Gamma$-centered $\veck$-point sampling defined with respect to the antiferromagnetic cell (2 NiO units).
(As an estimate of the remaining finite size error, the difference between the \gw@PBE~gaps for $4\times4\times4$ and $6\times6\times6$ $\veck$-meshes is only 0.1 eV).
We used the antiferromagnetic cell of 2 NiO units along the [111] direction as the impurity, corresponding to 78 impurity orbitals ($3s3p3d4s4p4d4f5s$ for Ni and $2s2p3s3p3d$ for O) and 72 bath orbitals in the DMFT impurity problem. As seen in Fig.~\ref{fig:nio}(a), \gw@PBE  severely
underestimates the band gap at 1.9 eV, even when using a spin symmetry broken PBE reference. We note that the quality of \GW~in this material
  is strongly dependent on the initial choice of Hamiltonian, and in practice improves through self-consistency, as has been seen in self-consistent quasiparticle \GW~calculations~\cite{Kotani2007}. Meanwhile, the valence spectrum of \gw@PBE does not
agree well with the experimental photoemission spectrum measured at a photon energy of 120 eV~\cite{Sawatzky1984}. We note that the experimental spectra do not report photoemission intensity in absolute units, so we rescale the spectra to make the main valence peaks of the experimental and GW+DMFT spectra approximately the same height, to facilitate comparison. \GW+DMFT, on the other hand, predicts a band gap of 4.0 eV and a magnetic moment of 1.69 $\mu_B$, both in very good agreement with the experimental values of 4.3 eV~\cite{Sawatzky1984} and 1.77-1.90 $\mu_B$~\cite{Fender1968,Cheetham83NiO}. More interestingly, our \GW+DMFT DOS captures the experimental two-peak structure of the valence spectrum around $-2$ and $-3$ eV. A detailed analysis of the spin-orbital-resolved local DOS in Fig.~\ref{fig:nio}(b) reveals that this two-peak structure  results from the splitting of
the majority and minority spin components of the Ni-$t_{2g}$ orbitals, and is a signature of the AFM phase, as it
  does not arise within the  paramagnetic phase~\cite{Kang2019}. Compared to the experimental valence spectrum measured at a lower photon energy (66 eV) that mainly probes O-$2p$ states, we find that our \GW+DMFT DOS agrees very well with the experimental peak at -5 eV, suggesting we have achieved a quantitative description of both local and non-local states in NiO. Our \GW+DMFT method also predicts a satellite peak around -8 eV, consistent with the photoemission spectrum in Ref.~\cite{VanElp1992}. We find this valence peak has a significant contribution from O-$2p$ orbitals, and a less substantial Ni-$3d$ weight. From the local DOS, we can also conclude that NiO is an insulator with mixed charge-transfer and Mott character,
with a valence band with contributions from Ni-$t_{2g}$, Ni-$e_g$ and O-$2p$, and a conduction band that is mainly of Ni-$e_g$ character.

\begin{figure}[hbt]
\centering
\includegraphics{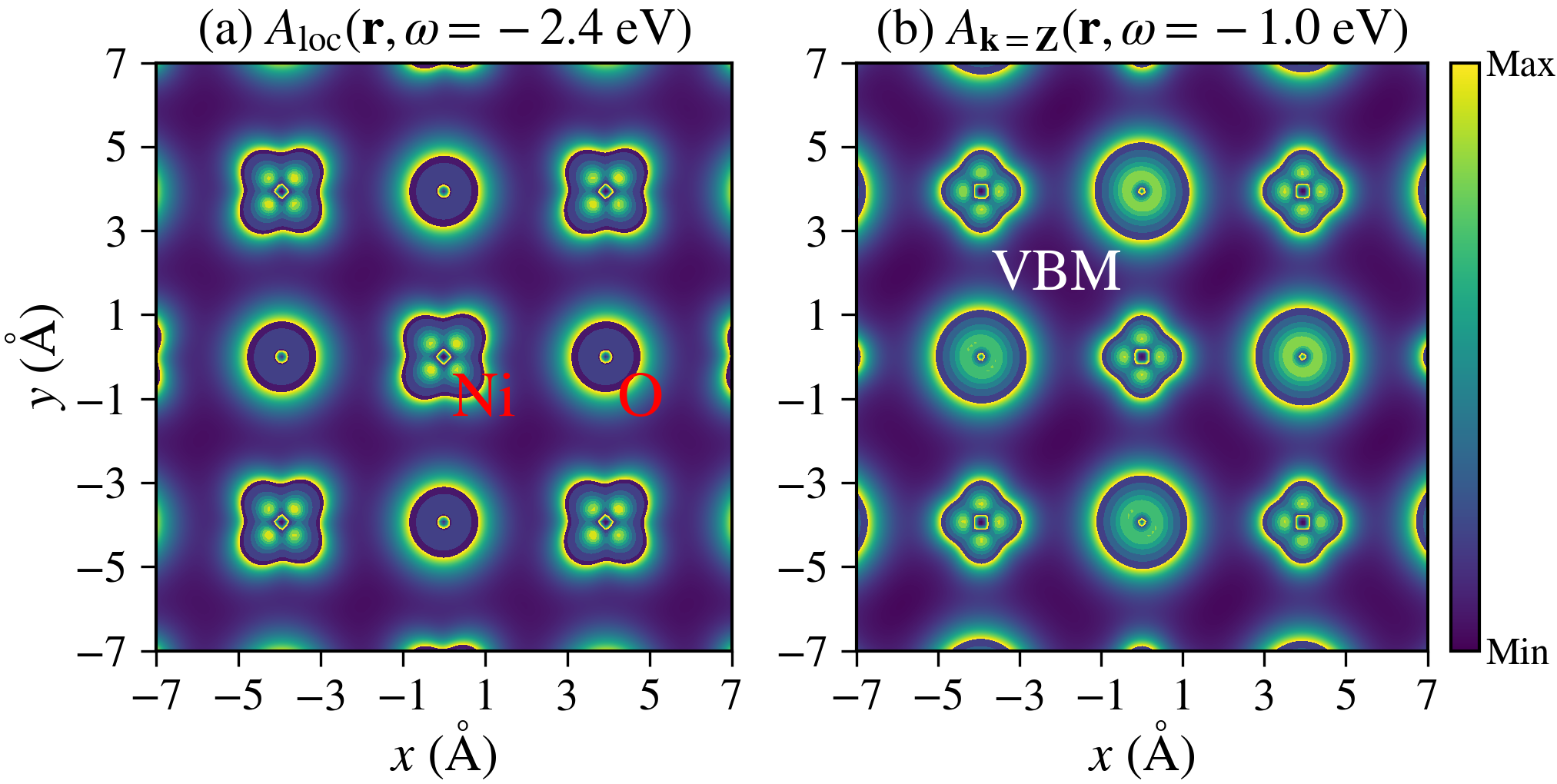}
\caption{Spatially-resolved DOS from \GW+DMFT (PBE reference) for NiO in the (001) plane. (a) Local DOS at $\omega=-2.4$ eV. (b) Momentum-resolved DOS at VBM energy $\omega=-1.0$ eV and $\mathrm{Z}=(0.5,0.5,0.5)$ point.}
\label{fig:space}
\end{figure}

In Figs.~\ref{fig:nio}(c) and \ref{fig:nio}(d), we further analyze the character of the conduction band minimum (CBM) and valence band maximum (VBM) in the Brillouin zone using the momentum-resolved DOS. We find that the lowest conduction band has strong Ni-$4s$ and O-$2s$ character at the $\Gamma$ point (CBM), which was not discussed in many earlier DMFT calculations~\cite{Kunes2007a,Leonov2016,Choi2016a} 
which focused on the Ni-$3d$ and O-$2p$ orbitals and thus did not include Ni-$4s$ (or O-$2s$) orbitals in the impurity (although see Ref.~\cite{Zhang2019} for a notable exception), unlike our full cell \GW+DMFT treatment.
At the Z point (VBM), we find that the highest valence band has significant O-$2p$ and Ni-$e_g$ contributions, with very little Ni-$t_{2g}$ character. This is very different from the local DOS, where the Ni-$t_{2g}$ has dominant weight in the valence bands. We confirm this by plotting the spatially-resolved DOS of NiO in the (001) plane in Fig.~\ref{fig:space}. We see that at the first valence peak and around the Ni atoms, the local spatial DOS has a Ni-$t_{2g}$ ($d_{xy}$) orbital shape, while the momentum-resolved spatial DOS (at the Z point) has a Ni-$e_g$ ($d_{x^2-y^2}$) orbital shape.
Further, the weight of the DOS around the O atoms in Fig.~\ref{fig:space}(b) is considerably larger than in Fig.~\ref{fig:space}(a). 

Since our impurity includes two NiO units, we can also look at correlations across the cells. We computed the spin-spin correlation function
for the Ni atoms within the impurity problem:
\begin{equation}
\sum_{{i\in \mathrm{Ni}_1, j\in \mathrm{Ni}_2}} \langle \mathbf{S}_i \cdot \mathbf{S}_j \rangle = \sum_{{i\in \mathrm{Ni}_1, j\in \mathrm{Ni}_2}} \sum_{a=x,y,z} \langle S_i^a S_j^a \rangle .
\end{equation}
We found $\langle \mathbf{S}_i \cdot \mathbf{S}_j \rangle$ between two Ni atoms to be -0.707.
Both $\langle S_i^x S_j^x \rangle$ and $\langle S_i^y S_j^y \rangle$ contribute almost zero spin correlation, and the uncorrelated value $\langle S_i^z \rangle \langle S_j^z \rangle$ is -0.710, suggesting that  quantum spin correlations are weak and NiO is close to a classical Ising magnet. This is consistent with experimental measurements of the critical behavior of the magnetic phase transition in NiO~\cite{Chatterji2009} and our previous \abinitio DMET study~\cite{Cui2020}. 

\begin{figure}[hbt]
\centering
\includegraphics{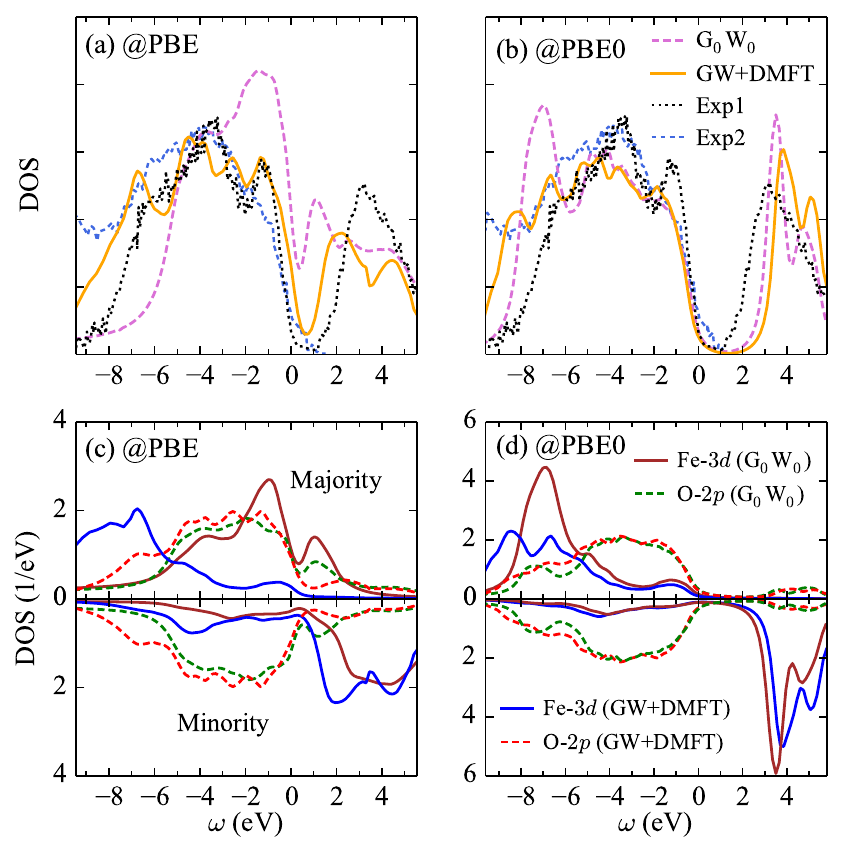}
\caption{Full cell \GW+DMFT results for \hematite~(AFM phase). (a)(b) Local DOS based on PBE and PBE0 references. The experimental spectra are taken from Ref.~\cite{Zimmermann1999} (exp1) and Ref.~\cite{Fujimori1986} (exp2). (c)(d) Orbital-resolved DOS corresponding to (a)(b). A broadening factor of 0.3 eV is used.}
\label{fig:fe2o3}
\end{figure}

We next turn to study a second strongly-correlated insulator, hematite (\hematite), in the AFM phase. We take the impurity to be the complete
AFM unit cell, including 2 \feo~units (Fig.~\ref{fig:illustration}), with a ``$+ - + -$'' type AFM ordering of the Fe spins.
Because of the large impurity size, we used a DZV-quality basis (GTH-DZV-MOLOPT-SR, $3s3p3d4s4p4d5s$ for Fe, $2s2p3s3p$ for O), leading to an impurity problem with 124 impurity and 48 bath orbitals. The small number of bath orbitals is due to the current numerical limitations of our CCSD solver. However, since our bath orbitals are only coupled to the valence impurity orbitals, and we aim to reproduce the hybridization only in a window near the Fermi level ($\pm 0.4$ a.u.), the bath discretization error is not too severe. Numerical tests (Supplemental Material) suggest that the finite bath discretization introduces an error of 0.05 $\mu_B$ in the Fe magnetic moment, while the band gap uncertainty is within 0.4 eV. The $3s3p$ orbitals of Fe were treated as frozen core orbitals (i.e., uncorrelated)
in the CCSD solver. The GTH-PBE pseudopotential and $4\times4\times4$ $\Gamma$-centered $\veck$-point sampling were employed.
As presented in Fig.~\ref{fig:fe2o3}(a), \gw@PBE severely underestimates the band gap at 0.5 eV,
compared to the experimental value of 2.6 eV~\cite{Zimmermann1999}. \gw~with the hybrid functional PBE0  slightly overestimates the gap (3.4 eV),
but the spectrum does not agree well with experiment, and in particular, the features of the \gw@PBE0 DOS are too sharp around -7 and 3.5 eV (Fig.~\ref{fig:fe2o3}(b)).

\GW+DMFT improves the \gw@PBE spectrum significantly, especially in the valence region, although the band gap (1.5 eV) is still too small. From
the orbital-resolved DOS in Fig.~\ref{fig:fe2o3}(c), we find that the main improvement comes from the spectral positions of the majority spin
component of the Fe-$3d$ orbitals and O-$2p$ orbitals. \gw@PBE mistakenly predicts the Fe-$3d$ valence spectrum to lie close to the Fermi level
and that \feo~has considerable Mott insulating character. However, \GW+DMFT shifts the majority-spin Fe-$3d$ DOS to lower energies, consistent with previous DFT+DMFT calculations~\cite{Kunes2009,Greenberg2018}. Because of this correction, \GW+DMFT obtains a more accurate Fe magnetic moment than PBE (4.23 $\mu_B$ compared to 3.71 $\mu_B$ with the experimental moment being 4.64 $\mu_B$~\cite{Coey1971}). While the \GW+DMFT DOS does not agree well with the ``exp1'' spectrum~\cite{Zimmermann1999} below -7 eV, we observe a much better agreement between \GW+DMFT and the ``exp2'' spectrum~\cite{Fujimori1986}, likely due to different experimental settings. We find that the valence band spectrum is dominated by O-$2p$  near
the Fermi level, indicating that \feo~is in fact a pure charge-transfer insulator, with almost no Mott insulating character. This is in contrast to
DFT+DMFT calculations~\cite{Kunes2009,Greenberg2018} that find a sizable Fe-$3d$ contribution to the valence band maximum.
We attribute this disagreement to the full cell \GW+DMFT treatment where both O-$2p$ orbitals
and Fe-$3d$ are treated on an equal footing at the 
impurity level, which thus allows for a more accurate balancing of their relative contributions to the spectral weight.

Starting from a PBE0 reference, \GW+DMFT finds a slightly larger band gap (3.9 eV) and magnetic moment (4.37 $\mu_B$) than \gw~(3.4 eV and 4.20 $\mu_B$). The overly sharp peaks of the \gw@PBE0 spectrum around $-7$ and 3.5 eV are corrected by \GW+DMFT, which broadens the Fe-$3d$ peaks as shown in Fig.~\ref{fig:fe2o3}(d). Comparing results between the PBE and PBE0 references, the severe reference dependence of spectral functions (especially Fe-$3d$ states) is largely reduced by \GW+DMFT. In summary, it appears we achieve a good description of the photoemission spectrum for \feo~within the full cell \GW+DMFT,
although a fully quantitative prediction of the band gap is not attained. Given that \gw~only provides a minor correction to the underlying DFT band gap in this system, the likely culprit is the insufficiency of the \gw~approximation
in describing the long-range interactions in \feo. 



We finally investigate a perovskite-type paramagnetic correlated metal \srmo~with two electrons ($4d^2$) occupying the Mo-$t_{2g}$ bands. We simulated the cubic structure of \srmo~using the GTH-PADE pseudopotential and GTH-DZVP-MOLOPT-SR basis set, with a $\Gamma$-centered $6\times6\times6$ $\veck$-mesh. To facilitate comparison with previous numerical studies, we used LDA as the underlying DFT functional. The \srmo~unit cell was taken as the DMFT impurity, corresponding to 71 impurity orbitals ($4d4f5s5p5d6s$ for Mo, $2s2p3s3p3d$ for O, $4d5s5p6s$ for Sr) and 117 bath orbitals. For paramagnetic metals, the quasiparticle peaks are expected to be sharp near the Fermi level, so we used a much smaller broadening ($\eta=0.02$ a.u.) when discretizing the hybridization on the real axis. We found this small broadening is crucial for avoiding non-causal behavior in the computed self-energies. 

\begin{figure}[hbt]
\centering
\includegraphics{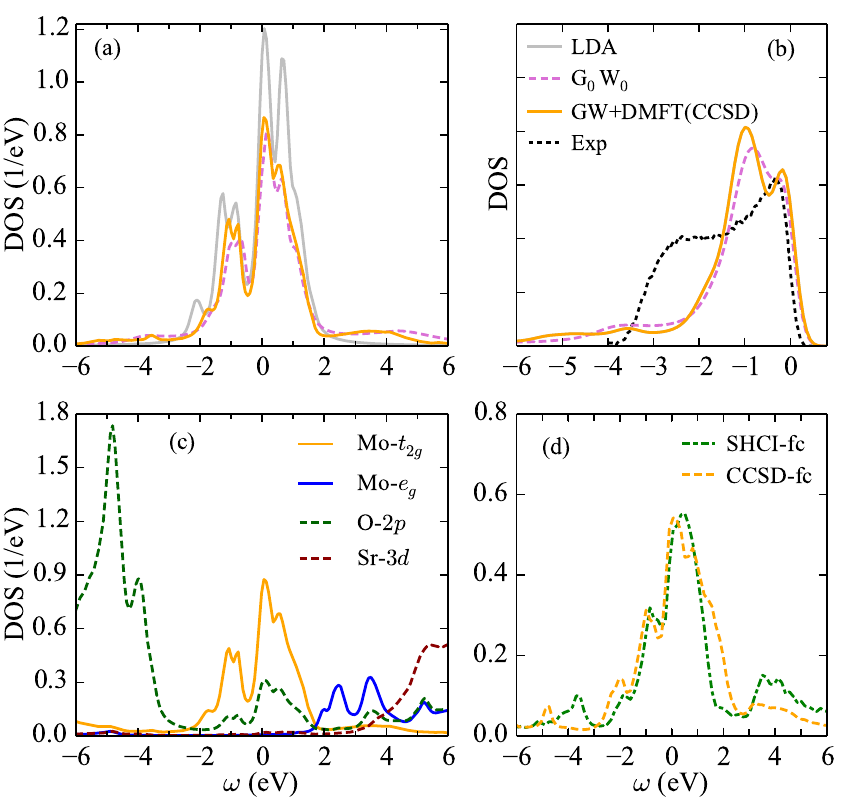}
\caption{Full cell \GW+DMFT results for paramagnetic \srmo~based on the LDA reference.  (a) Local DOS of Mo-$t_{2g}$ bands computed with a broadening of 0.15 eV. (b) Comparison of \gw~and \GW+DMFT with photoemission experimental data~\cite{Wadati2014}. A Fermi function of 298 K and a Gaussian filter of 0.2 eV are applied to the calculated DOS to match the experimental resolution. (c) Orbital-resolved DOS computed by \GW+DMFT (CCSD). (d) Comparison of \GW+DMFT DOS (Mo-$t_{2g}$ bands) using CCSD and SHCI solvers in a frozen-core correlated space.}
\label{fig:srmoo3}
\end{figure}

As the electronic structure of \srmo~near the Fermi level is dominated by Mo-$t_{2g}$ bands, we mainly discuss spectral functions of Mo-$t_{2g}$ bands in Fig.~\ref{fig:srmoo3}, although the orbital-resolved DOS is also included in Fig.~\ref{fig:srmoo3}(c). As shown in Fig.~\ref{fig:srmoo3}(a), \gw@LDA shows a small band narrowing compared to LDA, where the bandwidth is reduced from 3.9 eV (LDA) to 3.2 eV (\gw). \GW+DMFT with the CCSD solver predicts a slightly larger bandwidth of 3.5 eV than \gw~and similar quasiparticle bands near the Fermi level, which are clearly renormalized compared to LDA. In Ref.~\cite{Wadati2014}, an obvious narrowing of the quasiparticle bands compared to the LDA band structure calculation was not observed in the photoemission spectrum, which is incompatible with the measured enhancement of specific heat. This phenomenon was attributed to Hund's rule coupling, which induces strong quasiparticle renormalization even when the Hubbard interaction values are smaller than the overall bandwidth. Our \gw~and \GW+DMFT results are consistent with this observation. 

We further found a lower and upper sideband in the \gw@LDA spectrum, located around -3 and 4.5 eV. \GW+DMFT gives similar sidebands as \gw, but the position of the upper sideband is shifted to a lower value of 3.5 eV. As shown in Fig.~\ref{fig:srmoo3}(b), the experimental photoemission spectrum shows a substantial hump around -2.5 eV, which was not captured by the LDA+DMFT calculations. The authors in Ref.~\cite{Wadati2014} thus concluded this hump is likely a plasmon satellite, and proper treatment of long-range correlations is required to capture this effect. To compare with the photoemission spectrum, we applied a Fermi function of 298 K and a Gaussian filter of 0.2 eV on our calculated spectral functions to match the experimental resolution. We found that the quasiparticle bands near the Fermi level agree well with experiment in both \gw~and \GW+DMFT. However, although our \gw~and \GW+DMFT results predict sizable sidebands around -3 eV, the relative spectral weight of the sideband is too weak compared to experiment. This experimental hump is also unlikely to originate from other states in \srmo, as the \GW+DMFT O-$2p$ spectrum intensity is weak in the region of -3 to -2 eV as seen in Fig.~\ref{fig:srmoo3}(c). On one hand, this could indicate that larger impurity cells may be necessary in the full cell \GW+DMFT approach to
properly capture the strong plasmon intensity. However, we note that our results are in agreement with other \GW+DMFT calculations~\cite{Nilsson2017,Petocchi2020}, which may also suggest that the satellite intensity is overestimated in the experiment due to insufficiently high photon energies, or oxygen defects.

To demonstrate the use of a different quantum chemistry solver, and to benchmark the accuracy of our CCSD results for
\srmo~we also carried out calculations using the SHCI solver~\cite{Holmes2016,Sharma2017,Li2018}, within the full cell \GW+DMFT, as shown in Fig.~\ref{fig:srmoo3}(d). Because SHCI is much more computationally demanding than CCSD, we only performed a one-shot SHCI calculation on the impurity problem derived from the self-consistent \GW+DMFT solution using the CCSD solver, and used a variational threshold of $\epsilon=6\times10^{-4}$ Hartrees to select determinants. A
modified version of the Arrow code~\cite{Yao2020} was then used to compute the SHCI Green's function. To allow for a fair comparison, we restricted the
number of correlated electrons and orbitals in SHCI and CCSD to be identical (30 electrons in 131 orbitals), obtained
by freezing the lowest 57 occupied orbitals in the HF solution of the impurity problem.  Comparing the resulting \GW+DMFT Mo-$t_{2g}$ DOS, we find that the CCSD and SHCI spectrum agrees very well in \srmo, although SHCI predicts a slightly smaller bandwidth and stronger sideband intensity than CCSD.

\begin{figure}[hbt]
\centering
\subfigure{
\includegraphics{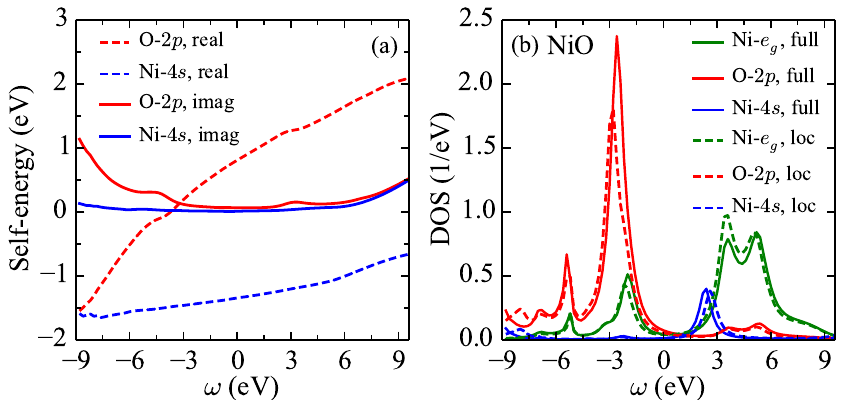}}
\vfill
\subfigure{
\includegraphics{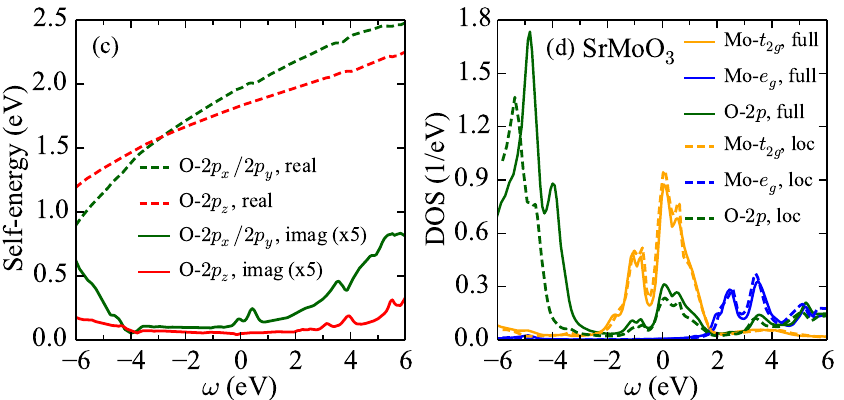}}
\vfill
\subfigure{
\includegraphics{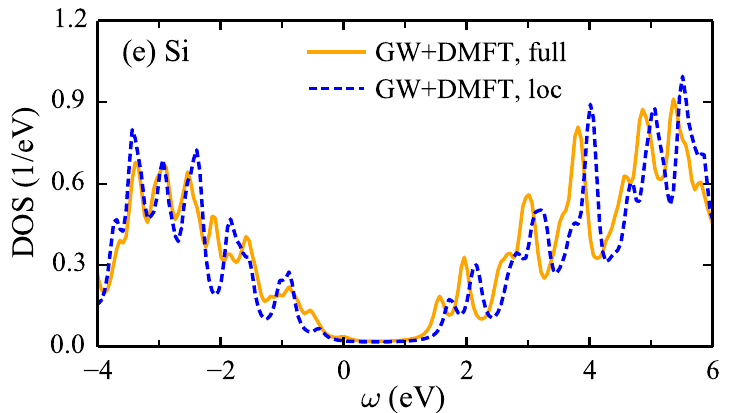}}%
\caption{Effects of non-$d$-orbital vertex corrections to the self-energy ($\vecSig^\mathrm{CC}-\vecSig^\GW$) on the full cell \GW+DMFT results. (a) DMFT self-energy corrections to the O-$2p$ and Ni-$4s$ orbitals in NiO. (b) \GW+DMFT $\veck$-resolved DOS of NiO at the  $\Gamma$ point,
    obtained by adding the self-energy correction to all impurity orbitals (``full'') or only the Ni-3d orbitals (``loc''). (c) DMFT self-energy corrections to O-$2p$ orbitals in \srmo. (d) \GW+DMFT local DOS of \srmo, obtained by adding the self-energy correction to all impurity orbitals (``full'') or only the Mo-$4d$ orbitals (``loc''). (e) \GW+DMFT local DOS of Si (HF reference), obtained by adding the self-energy correction to all impurity orbitals (``full'') or only the $3s3p$ orbitals within each Si atom (``loc'').}
\label{fig:offsig}
\end{figure}

Lastly, in Fig.~\ref{fig:offsig}, we perform an analysis to
understand the role of intermediate range interactions and vertex corrections that are now captured within 
our full cell \GW+DMFT method but omitted in the usual downfolding based \GW+DMFT treatments.
In the systems studied here, one way in which this shows up is in quantitative corrections to the physics of the non-$d$ orbitals of metal and non-metal atoms,
which are normally only treated at the DFT or \GW~level.
As shown in Fig.~\ref{fig:offsig}(a) and (c), there are significant vertex corrections to the self-energies of the O-$2p$ and Ni-$4s$ orbitals in NiO and O-$2p$ orbitals in \srmo~when using the full cell \GW+DMFT method. Such vertex corrections lead to non-trivial effects on the spectral functions of the non-$d$ orbitals. For example, the peak position and intensity of O-$2p$ DOS is clearly changed in both NiO and \srmo, when comparing the full and local (i.e., only Ni-$3d$ or Mo-$4d$) self-energy corrections ($\vecSig^\mathrm{CC}-\vecSig^\GW$). Furthermore, because some non-$d$ orbitals are strongly hybridized with the $d$ orbitals, correcting the non-$d$ orbitals also modifies the $d$-orbital DOS in NiO and \srmo. Meanwhile, in Fig.~\ref{fig:offsig}(e), we show that vertex corrections to the self-energies across two Si atoms within the impurity cell also lead to quantitative changes in the \GW+DMFT DOS of Si (based on the HF reference). In summary, our full cell \GW+DMFT provides vertex corrections to all orbitals, which is known to be necessary to achieve quantitative accuracy in many kinds of \abinitio simulations~\cite{Nilsson2017}.

\section{Conclusions}
In this work, we introduced a full cell \GW+DMFT formulation for the \abinitio simulation of correlated materials.
The primary strength of this approach is that it entirely avoids the problem of selecting a low-energy subspace,
and consequently the uncontrolled errors introduced either by downfolding the effective interactions within the subspace,
or via DFT double counting.
The resulting method is then fully diagrammatically controlled and can easily treat all interactions, and provides a framework to apply advanced quantum chemistry methods to study correlated solids. We showed that full cell \GW+DMFT
can be applied to systems using impurity cells of up to 10 atoms in calculations of the spectral properties of Si, NiO, \hematite~and \srmo,
obtaining for most quantities, results of good quantitative accuracy. 
By defining the impurity to comprise all orbitals in the AFM supercells of NiO and \hematite,
we also showed how the full cell approach can cleanly differentiate between different amounts
of charge-transfer and Mott insulating character and the orbital character around the gap, as both metal and non-metal orbitals enter into the impurity problem
on an equal footing.
Overall, our calculations demonstrate the potential of the full cell \GW+DMFT approach for studies of more complicated materials,
en route towards a fully predictive theory of correlated materials.

\begin{acknowledgments}
 This work was supported by the US Department of Energy via the
M$^2$QM EFRC under award no. de-sc0019330.
We thank Cyrus Umrigar and Yuan Yao for providing the SHCI Green's function code. TZ thank helpful discussions from Zhihao Cui, Xing Zhang and Timothy Berkelbach. Additional support was provided by the Simons Foundation via the Simons Collaboration on the Many Electron Problem, and via the Simons Investigatorship in Physics.
\end{acknowledgments}

%

\raggedbottom

\end{document}